# Eclipse time variations and the continued search for companions to short period eclipsing binary systems

George Faillace[1], David Pulley[1], John Mallett[1], Americo Watkins[1], Ian Sharp[1], Xinyu Mai[2]

[1] British Astronomical Association, Burlington House, Piccadilly, London W1J 0DU
[2] University of Iowa, Dept. of Physics and Astronomy, Iowa City, IA 52242, USA.
David Pulley: e-mail david@davidpulley.co.uk
George Faillace: e-mail gfaillace3@aol.com

*Abstract*

Eclipse time variations have been detected in a number of post common envelope binary systems consisting of a subdwarf B star or white dwarf primary star and cool M type or brown dwarf secondary. In this paper we consider circumbinary hypotheses of two sdB systems, HS 0705+6700 (also known as V470 Cam) and NSVS 14256825 and one white dwarf system, NN Ser. In addition, and for comparison purposes, we investigate the eclipse time variations of the W UMa system NSVS 01286630 with its stellar circumbinary companion. All four systems have claims of circumbinary objects with computed physical and orbital parameters. We report 108 new observations of minima for these four eclipsing systems observed between 2017 May and 2019 September and combining these with all published data, we investigate how well the published circumbinary object hypotheses fit with our new data. Our new data has shown departure from early predictions for three of the four systems, but it is premature to conclude that these results rule out the presence of circumbinary objects. There is also the possibility (but with no observational proof so far) of detecting close-in transiting circumbinary objects around these systems but these are likely to have periods of days rather than years.

*1. Introduction*

From observations of eclipse time variations (ETVs) many claims have been made for the detection of circumbinary objects orbiting subdwarf B (sdB) binary systems, members of the HW Vir family, and binary systems where the primary has evolved into a white dwarf. Typically, these systems have a very hot primary component with temperatures in excess of 30,000K and a secondary M dwarf or brown dwarf companion with temperatures of 3,500K or less. The separation between the two components is usually less than one solar radius causing the secondary to be heavily irradiated by the primary star giving rise to significant amounts of reflected energy from the secondary.

The structure of a HW Vir-type system, their compact structure, short periods and large temperature differences between their two components give rise to short and well-defined primary eclipses, allowing times of minima to be determined with high precision. These systems have undergone a post-common-envelope binary (PCEB) evolution, as described in Appendix 1, with many of these systems showing apparent periodic variations in their eclipse timings. An overview of these systems is provided by Zorotovic and Schreiber (2013) and Lohr et al. (2014).[1][2]

In this paper we consider three PCEB systems: HS0705+6700, NSVS 14256825, and NN Ser and one evolving to a W UMa-type of system: NSVS 01286630. Claims have been made for the presence of circumbinary objects, with calculated parameters, orbiting all four systems. Of all the putative objects discovered through ETVs, those around NN Ser were thought to be amongst the most compelling because of the high quality of the data and because its main sequence companion is a late M star which restricts the possibility of other causes for period changes, for example magnetic coupling. With our new observations we investigated the periodic variation in the position of the barycentre of these four systems to see if they fit with previous predictions of circumbinary orbits. While NN Ser and NSVS14256825 systems' planets are listed in various international databases, e.g. NASA Exoplanet Archive, neither of the other two systems have this level of recognition, calling into question their proposed circumbinary hypotheses.

While the more exciting explanation of attributing these period variations to the presence of planets or brown dwarfs orbiting these systems has been a popular consideration, other factors could also explain their cyclical behaviour (see Section 2, last paragraph). Other claims of circumbinary objects are of close eclipsing binary systems of the W UMa-type or binary systems evolving to a W UMa once their Roche lobes are filled, such as NSVS 01286630, the fourth system described herein.

We discuss the possibility that these other factors could explain the cyclical behaviour attributed in the literature to circumbinary objects for these four systems. We note that two of the systems (HS0705+6700 and NSVS 14256825) have been included in our recent study of seven short period eclipsing binary systems, Pulley et al. (2018).[3] As will be seen, our recent observations, which cover two additional seasons, support the conclusion of further deviations from previous predictions.

In this paper we present the ETVs exhibited by four eclipsing binary systems with somewhat different stages of binary evolution and we analyse the results in the context of circumbinary planet hypothesis. The analysis is preceded by a brief historical review of the hypotheses presented by earlier observers. The four systems studied are listed with their parameters in Tables 1 and 2.

| Object | RA (J2000) | Dec (J2000) | Period (d) | Distance (pc) | Mag. | New Observations Primary | Secondary | Period |
|---|---|---|---|---|---|---|---|---|
| HS 0705+6700 | 07 10 42.06 | +66 55 43.52 | 0.095646609 | 1001 | 14.60 (R) | 40 | 0 | 2017 Oct - 2019 Sep |
| NN Ser | 15 52 56.13 | +12 54 44.68 | 0.130080142 | 522 | 16.51 (V) | 16 | 0 | 2017 May - 2019 Aug |
| NSVS 01286630 | 18 47 08.58 | +78 42 29.34 | 0.383927870 | 324 | 13.09(V) | 20 | 2 | 2018 Jul - 2019 Sep |
| NSVS 14256825 | 20 20 00.48 | +04 37 56.49 | 0.110374168 | 838 | 13.34 (R) | 29 | 1 | 2017 Sep - 2019 Sep |

Table 1. Summary of the four objects observed between 2017 May and 2019 September with a total of 105 times of primary minima and 3 secondary minima.

| Object | $M_1$ ($M_o$) | $M_2$ ($M_o$) | $Teff_1$ (K) | $Teff2$ (K) | Sp. Type $M_1$ | Sp. Type $M_2$ | a ($R_o$) | i (deg) | Reference |
|---|---|---|---|---|---|---|---|---|---|
| HS 0705+6700 | 0.48 | 0.13 | 29600 | 2900 | sdB | M4/M5 | 0.81 | 84.4 | Drechsel et al. (2001) |
| NN Ser | 0.57 | 0.12 | 57000 | 2950 | WD | M4[1] | 0.95 | 84.6 | Brinkworth et al. (2008) |
| NSVS 01286630 | 0.68 | 0.72 | 4140 | 4290 | | | 0.72 | 89.0 | Coughlin & Shaw (2007) |
| NSVS 14256825 | 0.46 | 0.21 | 35250 | 3500 | sdB | M5/M7 | 0.74 | 82.5 | Kilkenny & Koen (2012) |

Table 2: Summary of key parameters of the binary systems observed. The spectral types for these systems are not clearly defined being indirectly determined from light curve parameters which themselves can be poorly constrained.
Notes: [1] Bours et al. (2016)

## *2. Observing Method and Data Reduction*

In Table 3 we list the telescopes used to make the new observations and in Table 4 (see Appendix 2 for Tables 3 and 4) we report 108 new observations between 2017 May and 2019 September of the four eclipsing binary systems. The effects of atmospheric extinctions were minimised by making all observations at altitudes greater than 40 degrees. All images were calibrated using dark, flat, and bias frames then analysed with Maxim DL or Astroart. [4][5] The source flux was determined with aperture photometry using a constant aperture for all images, and the radius scaled according to the full width at half maximum (FWHM). Variations in observing conditions were accounted for by determining the flux relative to an ensemble of comparison stars in the field of view. The apparent magnitude of the target was derived from the apparent magnitudes of the comparison stars and the average magnitude of the target calculated by the software. This was done as follows: The same comparison stars were used for each image. Using the average derived magnitude of the target star from each comparison star and the standard deviation of the average, the final value for the target was obtained for each frame. The magnitudes of the comparison stars were chosen appropriate to the filter being used. The comparison stars' catalogue magnitudes for the various filters were taken from the American Association of Variable Star Observers (AAVSO) Photometric All Sky Survey (APASS) catalogue [6], and were similar to the target magnitudes and, whenever possible, with similar colour indices to the target stars. Because the

APASS catalogue does not include the R pass band, in the few cases where observations were taken with the R filter a conversion formula recommended by AAVSO was used to transform the catalogue Sloan r' magnitudes to the corresponding R magnitudes. Whether observations were performed with or without filters, check stars were used to ensure that there was no variability in the reference star selected.

All of our new mid image timings used in this analysis were first converted to barycentric Julian date dynamical time (BJD_TBD) using the Ohio State University or the Astropy time utilities.[7][8] Computer clocks were synchronised with external atomic clocks during the imaging process. Times of minima were calculated using Kwee & van Woerden (1956) methodology and implemented with Peranso light curve analysis software.[9][10] Our new timings were combined with previously published times of minima and, where appropriate, the historic times were converted to BJD_TBD. Where a new linear or quadratic ephemeris was calculated only observed primary minima data were used. The difference between the observed and calculated times of minima were used to infer potential internal or external influences on the binary pair, for example (i) angular momentum loss through magnetic braking or the emission of gravitational waves; (ii) angular momentum redistribution through Applegate-type mechanisms; (iii) the apparent changing of the binary period through the presence of a circumbinary object; or (iv) apsidal motion, see for example Brinkworth et al. (2006); Bours et al. (2016) and references therein.[11][12]

## 3 Analysis of eclipse timings

### 3.1 HS0705+6700 (V470 Cam)

#### 3.1.1 Background

HS0705+6700 is a 14.6 magnitude sdB star, first identified in the Hamburg Schmidt Quasar Survey. Subsequent observations by Drechsel et al. (2001) confirmed this sdB object as a member of a short period, 2.3hr, PCEB system.[13] Light curve analysis indicated that the secondary companion was a low mass red dwarf star with mass and radius of $0.13M_0$ and $0.19R_0$ respectively. Niarchos et al. (2003) confirmed this structure but observations by Qian et al. (2009) suggested that the binary period had a superimposed cyclical component with a period 7.15yrs and Light Travel Time (LTT) amplitude of 92.4s.[14][15] The LTT effect in eclipsing binaries occurs because the distance to the observer varies due to the reflex motion of circumbinary companions moving around the barycentre of multiple systems. Qian's analysis ruled out period change due to apsidal motion, magnetic coupling (Applegate effect) and angular momentum loss and they concluded that the most likely cause of the observed period change was the presence of a brown dwarf circumbinary companion of mass $39.5M_J$. Further observations and analysis by Qian et al. (2010) and (2013),[16][17] Camurdan et al. (2012) and Beuermann et al. (2012) strengthen this prediction by providing some 15 years of data spanning ~1.7 circumbinary periods.[18][19] Qian's revised analysis suggested a brown dwarf circumbinary companion of mass $32M_J$ and with a period of 8.87yr.

Although observations by Pulley et al. (2015) initially confirmed these findings they noted in their addendum that data post 2015 February indicated a departure from the proposed circumbinary model.[20] This departure was confirmed by Pulley et al. (2018) bringing into question the circumbinary brown dwarf prediction.[3].

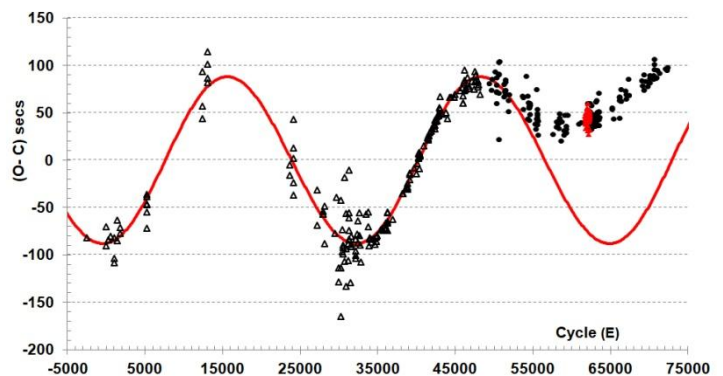

Fig. 1: HS0705+6700 (O - C) residuals with the 8.55yr circumbinary brown dwarf companion shown as the red line. Open triangles are known historical data; red triangles are Bosenberger's data corrected for mid exposure time; our new data is shown as solid black circles.

Further data and analysis by Bogensberger et al. (2017), using a linear ephemeris, resurrected the circumbinary brown dwarf prediction but with a significantly longer period of 11.8 years and orbital eccentricity of 0.38. [21]

*3.1.2 New observations and ephemeris*

We provide 40 new times of minima taken from observations made between 2017 October and 2019 September. As shown by Qian et al. (2013), the quadratic ephemeris provides a better fit to the data than the corresponding linear ephemeris through to 2015 March.[17] Utilising the quadratic ephemeris of Pulley et al. (2018), [3] Eq. 1 below, the plot of (O-C) residuals is shown in Fig.1, where the red line represents the predicted (O – C) residuals incorporating the circumbinary companion.

$$T_{min;BJD} = 2451822.76155(5) + 0.095646609(4) *E + 5.5(9)* 10^{-13} *E^2 + \tau \qquad (1)$$

$\tau$ is the cyclical light travel time effect of a putative circumbinary object given by:

$$\tau = \frac{a_{12}\sin i}{c}\left[(1-e^2)\frac{\sin(v+w)}{1+e\cos v} + e\sin w\right] \qquad (2)$$

and $(a_{12}\sin i)/c$ is the LTT amplitude of 88.1s; $e$ is the circumbinary component orbital eccentricity of 0.03; $w$, the angle of periastron is 0.119 rads and $v$ is the true anomaly and determined from the time at periastron of 2449484.0 days and circumbinary period of 8.55 yrs.

Our new data (E > 65115) is not consistent with the findings of Bogensberger et al. but is consistent with our earlier findings confirming the departure from the pre 2015 circumbinary prediction of Qian, Camurdan, Beuermann and Bogensberger. Using the post 2015 March data only, we compute a new linear ephemeris, Eq. 3:

$$T_{min, BJD} = 2451822.75657(16) + 0.095646732(3) *E + \tau \qquad (3)$$

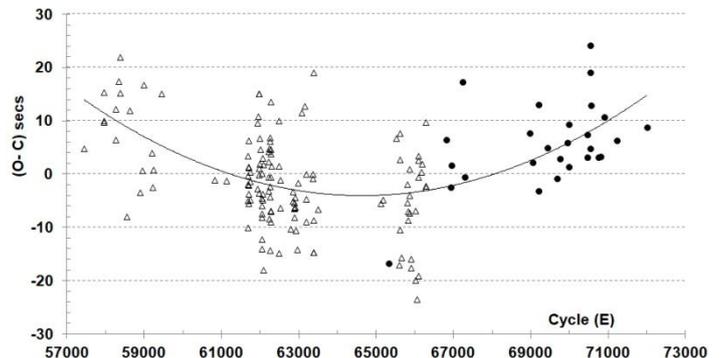

The (O – C) residuals, based on Eq.3, are shown in Fig. 2 where there is seen to be a quadratic trend. A comparison of the linear and quadratic fits for this latter data shows that the reduced $\chi^2$ quadratic fit at 1.62 is lower than the reduced $\chi^2$ linear fit of 2.75. The quadratic ephemeris for this recent data yields a quadratic coefficient at 3.98 x $10^{-12}$ days, nearly order of magnitude larger that the quadratic term of the pre-2015 data which may reflect the possible continuing presence of a third body. However, application of the Mann-Whitney U test shows no strong preference for either of these two models at the p(0.05) level.

Fig. 2: HS0705+6700 (O - C) residuals post 2015 March and derived from the new linear ephemeris, Eq. 3, indicating the presence of a quadratic component but no strong indication of the earlier cyclical component of amplitude of ~85s (see Fig. 1). Historic data is shown with open triangles and our latest data as solid circles.

*3.2 NSVS 14256825 (V1828 Aql)*

*3.2.1 Background*

Since its discovery there have been many predictions for circumbinary objects orbiting NSVS 14256825. This binary system was first discovered as a 13.2 magnitude variable star during the Northern Sky

Variability Survey (NSVS). Subsequently Wils et al. (2007) identified it as a short period, ~2.6 hr, eclipsing binary with sdOB primary and a cool M dwarf or brown dwarf secondary. [22]

Observations by Qian et al. (2010) indicated a cyclical change in the binary period and attributed it to a possible third body.[16] Kilkenny and Koen (2012) published nine times of minima noting the binary period was rapidly increasing.[23] Beuermann et al. (2012) recorded 27 new times of minima and included a further five times extracted from the ASAS (All Sky Automated Survey) and NSVS databases using phase folding techniques.[19] These five data points had significantly larger uncertainties, of the order of 50s, but enabled the timeline to be extended back almost 8 years to 1999. They also noted that post 2009 the binary period increased significantly suggesting the presence of a poorly constrained circumbinary object of mass ~12$M_J$ and period of some 20yrs.

Almeida et al. (2013) performed a new circumbinary analysis while presenting ten new times of minima.[24] They interpreted the binary period variations as the result of LTT effects introduced by two circumbinary planets with orbital periods of 3.5 yr. and 6.9 yr. and masses 3 $M_J$ and 8 $M_J$, respectively. Subsequently Hinse et al. (2014) suggested that a "local minima" had been found and that a longer observational timeline was needed.[25] Similarly, an analysis by Wittenmeyer et al. (2013) showed that the proposed two planets are dynamically unstable with a projected lifetime of <1 Myr and substantially shorter than the age of this system.[26]

A third circumbinary model was put forward by Nasiroglu et al. (2017) who combined a further 83 new times of minima spanning 2009 August to 2016 November with existing data but excluded data from ASAS, NSVS and SuperWASP due to their large uncertainties.[27] Their analysis suggested a possible brown dwarf circumbinary companion with a minimum mass of ~15$M_J$ and period of ~10 yr. Their results, and a further 19 times of minima from Pulley et al. (2018),[3] confirmed that the Almeida two planet model failed to correctly predict eclipse times beyond 2013 March.

A further 84 times of minima published by Zhu et al. (2019) showed a deviation from the Nasiroglu circumbinary brown dwarf model.[28] With their new observations, Zhu revised the circumbinary prediction with a brown dwarf having a minimum mass of 14.15$M_J$ and orbital period of 8.83 years.

### 3.2.2 New observations

We report a further 30 times of minima observed between 2017 September and 2019 September and, excluding ASAS and NSVS data points, we compute a new ephemeris, Eq.4:

$$T_{min\ BJD} = 2454274.20921(4) + 0.110374105(2) * E + \tau \qquad (4)$$

where $\tau$ is the LTT contribution from the third body. We compute the LTT parameters, $\tau$, from Eq. 2. For the purposes of this paper we have followed the Nasiroglu and Zhu approach and omitted the five sky survey data points, computing the parameters for a single circumbinary model. Parameters for this model are listed in Table 5 alongside earlier published models. The plot of (O – C) residuals for our new model are shown in Fig. 3 where the early ASAS and NSVS datasets, together with their uncertainties, have been shown for completeness.

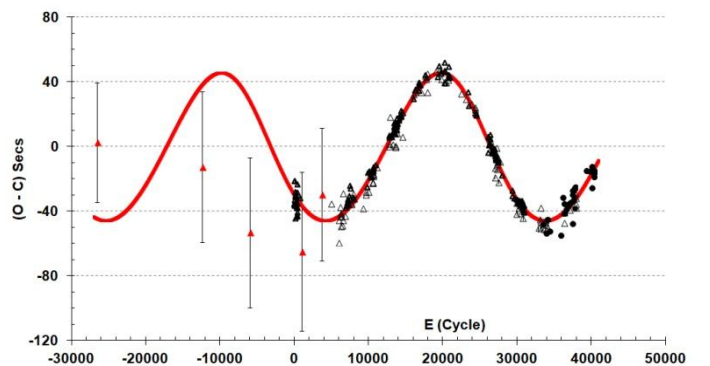

Fig. 3: (O - C) residuals for NSVS 14256825 based on our revised ephemeris, Eq. 4, and single circumbinary object hypothesis. Open triangles indicate historical data; solid black circles are our new data; red triangles are Beuermann's ASAS and NSVS phase folded data. The red curve represents the predicted (O - C) residuals based on the proposed third body.

|  |  | Beuermann 2012 | Almeida 2013 | | Nasiroglu 2017 | Zhu 2019 | This paper |
|---|---|---|---|---|---|---|---|
| *Binary parameter* | | | | | | | |
| Binary epoch (+240000) | BJD | 54274.208923(4) | 54274.20874(4) | | 55793.84005(3) | 54274.20921(1) | 54274.20921(4) |
| Binary period | days | 0.1103741324(3) | 0.1103741681(5) | | 0.110374099(3) | 0.1103741030(5) | 0.110374105(2) |
| *Third body* | | | Body 1 | Body 2 | | | |
| *Star survey data included* | | Yes | Yes | | No | No | No |
| LTT Amplitude | secs | 59 | 85(3) | 5.0(3) | 48.10(1) | 46.3(4) | 46.0(5) |
| Eccentricity | | 0.5 | 0.52(8) | 0.00(+8) | 0.175(12) | 0.12(2) | 0.11(10) |
| Period | years | 20 | 6.86(45) | 3.49(38) | 10.96(41) | 8.83(6) | 8.94(51) |
| Periastron passage | BJD | 2454836 | 2456643(110) | 2455515(95) | 7938.5(204) | 2456816(94) | 2456825(48) |
| Longitude of periastron | rads | 4.57 | 1.71(15) | 0.19(14) | 1.57(24) | 2.33(18) | 2.37(10) |
| Mass (i = 90$^0$) | M$_J$ | 12 | 8.0(15) | 2.10(4) | 14.75(13) | 14.15(0.16) | 14.53 |

Table 5: Comparison of the circumbinary models for NSVS 14256825. Recent data from Zhu et al. (2019) and this paper show a departure from the Nasiroglu et al. (2017) model and reflected in the smaller orbital eccentricity and shorter third body period.
Note 1: Parameters for Beuermann et al. (2012) are poorly constrained
Note 2: Uncertainties specified for Nasiroglu et al. (2017) are asymmetric. Those above are mean values
Note 3: Almeida used two of the five ASAS/NSVS data points ~ cycle 1018 and 3737

### *3.3 NN Ser*

#### *3.3.1 Background*

NN Ser is a 16.5 magnitude short period, ~3.1 hr, binary system with a white dwarf primary and M dwarf secondary sharing many evolution similarities to that of the sdB binary NSVS 14256825. It is the only WD short period binary to show strong evidence to support the presence of circumbinary objects. NN Ser was first investigated in 1989 by Haefner who identified it as a pre-cataclysmic binary with a deep primary eclipse (>4.0 mag) with strong reflection effects from the close by M dwarf secondary.[29] System parameters were first determined by Woods & Marsh (1991) and refined by Catalan et al. in 1994.[30][31] Further times of minima were obtained by Haefner et al. in 2004.[32]

Variability in the period of this system was first noted by Brinkworth et al. in 2006 when they reported 13 new times of minima, whilst extending the observational baseline to 15 years.[11] Their analysis ruled out many possible causes of the period change, e.g. apsidal motion, gravitational waves and magnetic Applegate effects, preferring either angular momentum loss through magnetic breaking or the possibility of a low mass circumbinary companion. However, their parameters for a putative third body were poorly constrained. Qian et al. (2009) added five new times of minima and suggested that the data indicated a cyclical change in the binary period with a superimposed long term period decrease.[33] They attributed the cyclical effect to the presence of a circumbinary planet mass <14.7M$_J$ and assigning the underlying period decrease to magnetic breaking.

Beuermann et al. (2010) provided new eclipse times and suggested the period variation could be attributed to two circumbinary objects with minimum masses of 6.9M$_J$ and 2.2M$_J$ and periods in a 2:1 resonance of 15.5yr and 7.7yr respectively.[34] Beuermann et al. (2013) also published a further 69 times of minimum and reconfirmed their earlier two planet model but with marginally refined orbital parameters.[35] Marsh et al. (2014) provided a further 25 times of minima confirming the presence of the two planets with orbital parameters very similar to Beuermann's 2013 prediction.[36]

Whilst the circumbinary planet hypothesis was the most favoured explanation for the observed period variations, Parsons et al. (2013) explored apsidal motion as the possible cause.[37] They observed 16 secondary eclipses of NN Ser and determined the eccentricity of the binary orbit to be less than 0.001 thus ruling out apsidal precession as a cause of the period variations. A long term study of 67 WD short period binaries was reported by Bours et al. (2016) and identified the secondary companion to NN Ser as an M dwarf of spectral type M4.[12] They added a further 10 times of minima and confirmed that their new data fitted a two planet model but noting that to get the best fit they had to add an additional quadratic term to the ephemeris. This new term causes the period to increase with the most probable explanation being a distant third body. Hardy et al. (2016) obtained more observations from Atacama Large Millimetre Array (ALMA) detecting thermal emissions from a dust disc of mass ~ 0.8 $M_{earth}$ and postulated its origin to be from common envelope material not expelled from the system.[38] They argued that, whilst not confirming the existence of planets, this added to the possibility of formation of the so-called ''second generation'' planets.

### 3.3.2  New observations and ephemeris

We report 16 new times of minima of NN Ser observed between 2017 May and 2019 August. The light curve for NN Ser's primary eclipse, Fig. 4, shows the compact nature of the WD primary which gives rise to a deep primary minimum with steep ingress and egress. Our times of minima have been calculated by fitting straight lines to the ingress and the egress portions of the light curve and determining the mid-point between these two lines. Uncertainties of less than 3s were achieved with 2-meter aperture telescopes. For smaller apertures uncertainties were substantially higher, see Table 4.

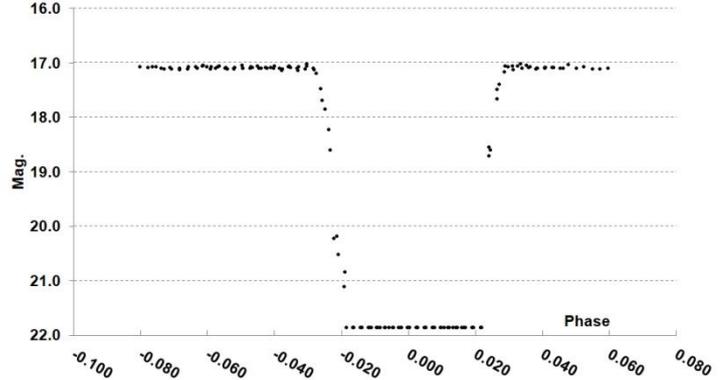

Fig. 4: The light curve of NN Ser's primary minimum showing the steep ingress and egress as would be expected from the eclipse of a compact white dwarf primary star. The out of eclipse rising and falling shoulders from the secondary reflected light are also present in the NN Ser light curve.

With our new data, together with all known published data, and using Beuermann et al. (2013) ephemeris:[35]

$$T_{min\ BJD} = 2447344.524368(7) + 0.13008014203(3) * E + \tau \qquad (5)$$

where $\tau$ is the circumbinary LTT parameters, the resulting (O – C) residuals are shown in Fig. 5. We first noted a departure from the two circumbinary planet model in mid-2017 (E~ 80,000). With our recent data we can confirm that the original Beuermann, Marsh and Bours predicted two planet circumbinary model needs to be reviewed as more data is collected.

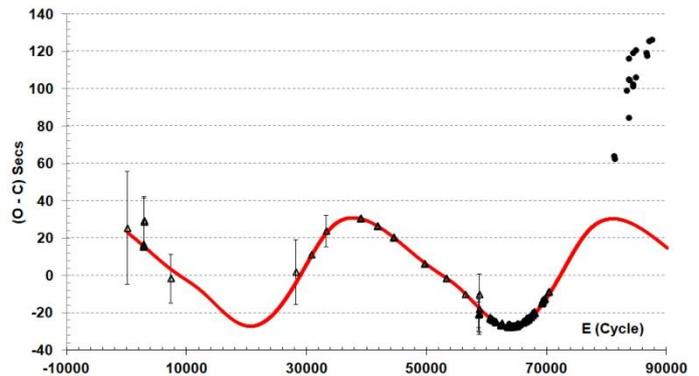

Fig. 5: NN Ser (O - C) residuals with linear ephemeris removed. Historical data is shown with open triangles and solid points are our new data. The red line is the two planet hypothesis of Beuermann et al. (2013). Observations depart from predictions at E~80,000 corresponding to mid 2017.

### 3.4 NSVS 01286630

### 3.4.1  Background

NSVS 01286630 (NSVS 1135262 in SIMBAD) is a detached eclipsing binary with a short orbital period of 9.2hr with deep symmetric primary and secondary eclipses first identified as a low mass binary (LMB) by Shaw & Lopez-Morales in (2007).[39]

Coughlin & Shaw (2007) observed light curves and derived stellar parameters for this system determining its principal properties as $M_1$ =0.68$M_o$, $R_1$=0.081$R_o$, $T_1$=4140K and $M_2$=0.72 $M_o$, $R_2$ =0.87$R_o$, $T_2$=4290K.[40] They noted that LMBs had a high level of star spot activity and, to achieve the best light curve fit, they modelled the system with two star spots on the hotter secondary component.

Wolf et al. (2016) added 94 times of minima, observed between 2005 November and 2016 May, and identified a cyclical variation in the binary period which they attributed to a LTT effect.[41] From this they deduced the presence of a 0.103 stellar mass third body orbiting with a period of 3.62yr. Zhang et al. (2018) observed five times of minima between 2010 November and 2011 June deriving a similar ephemeris and third body parameters to that proposed by Wolf. [42] Both Wolf et al. and Zhang et al. considered the possibility of a magnetic Applegate type mechanism driving the LTT effect but discarded this in favour of a circumbinary object.

*3.4.2 New observations and ephemeris*

To investigate whether the circumbinary model of Wolf and Zhang continues to hold beyond 2015 November, we present 22 new times of primary minima observed between 2018 July and 2019 September. We have adopted the ephemeris of Zhang et al. (2018) first converting time from HJD to BJD:

$$T_{min\ BJD} = 2454272.753960 + 0.3839278700 * E + \tau \qquad (6)$$

where the light travel time effect, $\tau$, is given by Eq. 2 and $(a_{12}\sin i)/c$ is the light travel time amplitude of 83.8s, the tertiary component orbit eccentricity, e, of 0.08. Application of this ephemeris with our new data is shown in Fig. 6 strongly suggesting that the circumbinary models of both Wolf and Zhang may need modifying over this extended timeline. More data over the coming months is required to clarify this potential anomaly.

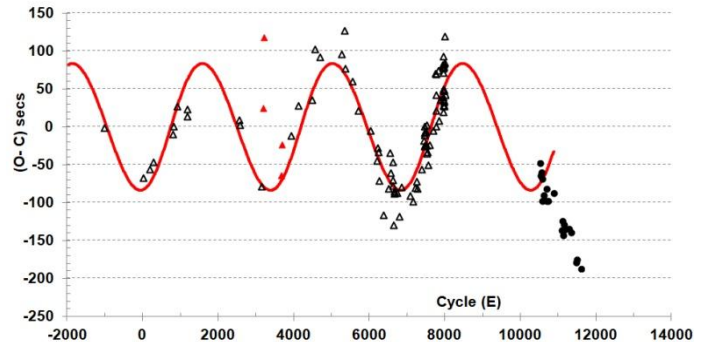

Fig. 6: The plot of (O - C) residuals for NSVS 01286630 adopting the ephemeris and circumbinary properties from Zhang et al. (2018). Our new data is shown as solid black circles; all historical data shown as open triangles with the Zhang data as red triangles. Zhang's fifth point at (3804, -334.99) is out of range of this chart.

*4.0 Discussion*

Historically, the approach adopted to determine the presence of circumbinary objects around sdB eclipsing binaries has been to first identify a cyclical behaviour within the eclipse timing measurements and then to note all possible causes. The improbable causes are then eliminated and whatever remains is assumed to provide the most likely explanation of the observed cyclical behaviour. In the case of sdB PCEB systems the remaining cause has invariably been attributed to a circumbinary companion. Whilst this reasoning is sound, it does contain a number of underlying assumptions including (i) having a knowledge of all possible causes of cyclical ETVs (ii) having a complete understanding of the mechanisms behind these causes and (iii) the cyclical behaviour is stable over at least several periods of each circumbinary object.

However, the complexity of these systems may undermine some of these assumptions. Whilst these systems are described simplistically as detached, i.e. there is no mass transfer between them, the reality, as modelling has shown, may be far more complex. The "surfaces" of the binary pair are separated by typically less than a solar radius and with the effective temperature of the secondary companion being of the order of 3000K, it is heavily irradiated by the primary which is significantly hotter at circa 30000K. Whilst there is no mass transfer, their energy interaction is likely to be complex and a modified Applegate type mechanism may be significant. It is frequently noted from light curve solutions of these systems that an improved fit can be found when the bolometric albedo of the secondary is set to a nonphysical value of greater than unity, see for example Drechsel et al. (2001).[13]

There too remains the open question of how circumbinary planets form around these systems and whether they are first- or second-generation objects, i.e. were they formed before or after the common envelope ejection, or possibly a combination of both, see for example Schleicher & Dreizler (2014). [43]

We note in three of the four systems we have investigated (HS0705+6700, NN Ser and NSVS 01286630) there is a significant visual departure from the historic (O − C), possibly indicating a change within the binary system. These departures indicate a small increment of a few tens of milliseconds in the PCEB period and could mask the presence of circumbinary objects. This is particularly so for HS0705+6700 where we have gathered four years of data since the observed change reflecting a binary period increase of some 11ms. Whilst seemingly small it is significant in comparison with the 8300s period of the binary system. If a similar magnitude change were to occur between the Earth and Sun, we would see the Earth year increase by 40s over a period of a few months.

If there was a circumbinary object present prior to this binary period increase, and there are 1.7 cycles of data to suggest this, the question remains as to whether this object is still present. Our analysis of the last four years of data suggests that, so far, there is no strong indication for the presence of a cyclical signal attributable to a circumbinary object.

NN Ser shows a similar trend to HS0705+6700 with a small period increase noted from mid-2017. Unlike its sdB counterpart, this system does not, as yet, give any strong indication of a cyclical or quadratic component within the new data. However, with it's out of eclipse magnitude of 16.5, NN Ser does require the use of large optics to acquire high precision data.

NSVS 01286630, a W UMa type object which has not progressed to the PCE stage, has been included to provide a contrast with the other three short period PCEB systems reported here. Whilst this object has shown a binary period decrease of 110ms, there is again no clear indication of a circumbinary or quadratic component in the new data. However, with little more than one year of new data, a longer timeline is necessary.

In contrast to the above three systems, the sdB NSVS 14256825 does show a cyclical 8.9-year period with data spanning the past twelve years. However, as new data has been acquired, each previous proposed circumbinary model has failed. The validity of our new model will only be confirmed when observations over the coming five or six years provide two full cycles of consistent data. We note that all recent analyses of this system have ignored the early phase folded light curve data of Beuermann et al. [19]

*5 Conclusions*

Whilst it is well known that observations of eclipse time variations can be used to detect circumbinary objects, the considerations outlined in Section 4 can make it difficult to interpret ETVs particularly if timelines are short, as they frequently are, in comparison with the longest circumbinary period. For acceptable confidence levels, minimum observational timelines of twice the longest circumbinary period are suggested. These difficulties are reflected in the NASA Exoplanet Archive where there are only 11 exoplanets listed using the ETV methodology out of a total of some 4000 listed exoplanets. There are also instances where claims have been refuted and planets removed, e.g. HW Vir.

Whilst our new data and analysis conflicts with earlier interpretations, and in most cases does not necessarily confirm the presence of stable circumbinary objects around these systems, it may also indicate that post common envelope binary systems are going through complex transitory phases. Our new data has shown departure from early predictions for three of the four systems, but it is premature to conclude that these results rule out the presence of circumbinary objects.

Of the systems we have investigated in this and our previous papers, only three have confirmed circumbinary planets included in the NASA Exoplanet Archive, NY Vir b, NSVS 14256825 b and NN Ser c & d. It is noted that the longer, ~25-year period companion to NY Vir, computed by both Lee and Song, has not been included in the NASA database, and NSVS14256825 has, so far, only 1.3 cycles of confirmed data. Similarly, only one of some 40 WD PCEBs, NN Ser, has identified circumbinary planets. Whilst NN Ser's two circumbinary planets have been listed on the NASA database our new results now raise questions on their presence/parameters.

Finally, we reiterate additional observations over a longer timeline are required to increase confidence in the many circumbinary claims of post common envelope binary systems that have been made over the past decade. Indeed, this increased timeline may show further complex changes in these ETVs.

*Acknowledgements*

This work makes use of observations from the Las Cumbres Observatory network of telescopes and of the APASS database maintained on the AAVSO website. We are particularly indebted to Dr. S von Harrach for his important contributions in enhancing this paper and Peter Starr of Warrumbungle Observatory for providing many images. We would also like to thank Dr Y-L Zhu and Dr Bin Zhang of the Chinese Academy of Sciences and Dr Leonardo Almeida of the Federal University of Rio Grande do Norte Natal, Brazil, who provided background on their research activities in support of our work. Finally, we want to thank the two referees for making important suggestions which improved the readability of the paper.

***References and notes***

1. Zorotovic M. & Schreiber M. R. 2013, A&A, 549, A95
2. Lohr M. E., Norton A. J., Anderson D. R., et al. 2014, A&A, 566, A128
3. Pulley D., Faillace G., Smith, D., Watkins A., von Harrach S. 2018, A&A, 611, A48
4. http://diffractionlimited.com/product/maxim-dl/
5. http://www.msb-astroart.com/
6. APASS: www.aavso.org/apass (2015)
7. http://astroutils.astronomy.ohio-state.edu/time
8. https://www.astropy.org/
9. Kwee K. & van Woerden H. 1956, Bull. Astron. Inst. Neth., 12, 327
10. http://www.cbabelgium.com/peranso/
11. Brinkworth C., Marsh T., Dhillon V., & Knigge C. 2006, MNRAS, 365, 287
12. Bours M. C., Marsh T., Parsons S., et al. 2016, MNRAS, 460, 3873
13. Drechsel, H., Heber, U., Napiwotzki, R., et al. 2001, A&A, 379, 893
14. Niarchos, P., Gazeas, K., & Manimanis, V. ASP Conference Series, Vol. 292, 2003
15. Qian, S.-B., Zhu, L.-Y., Zola, S., et al. 2009, ApJ, 695, L163
16. Qian, S.-B, Zhu, L., Liu, L., et al. 2010, Astrophys. Space Sci., 329, 113
17. Qian, S.-B., Shi, G., Zola, S., et al. 2013, MNRAS, 436, 1408
18. Çamurdan, C. M., Zengin Çamurdan, D., & Ibanoglu, C. 2012, New Astron., 17, 325
19. Beuermann, K., Breitenstein, P., Debski, B., et al. 2012, A&A, 540, A8
20. Pulley, D., Faillace, G., Smith, D., Watkins, A., & Owen, C. 2015, J. Brit. Astron. Assoc., 125, 5
21. Bogensberger, D., Clarke, F., Lynas-Gray, A. 2017, Open Aston., 26, 134
22. Wils, P., di Scala, G., & Otero, S. A. 2007, IBVS, 5800, 1
23. Kilkenny, D., & Koen, C. 2012, MNRAS, 421, 3238
24. Almeida, L. A., Jablonski, F., & Rodrigues, C. V. 2013, ApJ, 766, 11
25. Hinse, T. C., Lee, J.W., Go´zdziewski, K., Horner, J., &Wittenmyer, R. A. 2014, MNRAS, 438, 307
26. Wittenmyer, R. A., Horner, J., & Marshall, J. P. 2013, MNRAS, 431, 2150
27. Nasiroglu, I., Go´zdziewski, K., Słowikowska, A., et al. 2017, AJ, 153, 137
28. Zhu L.-Y. et al. 2019, arXiv :1904.11664v3, ApJ, 116,211
29. Haefner R., 1989, A&A, 213, L15
30. Woods J.H., Marsh T.R., 1991, ApJ, 381, 551
31. Catalan M.S., Davey S.C., et al. 1994, MNRAS, 269, 879
32. Haefner R., Fiedler, A., Butler, K., Barwig, H., 2004, A&A, 428, 181
33. Qian S.B., Dai Z.B., Liao W.P., et al., 2009, ApJ. 706, L96
34. Beuermann, K. et al. 2010, A&A, 521, L60
35. Beuermann, K. et al. 2013, A&A 555, A133
36. Marsh T.R., et al. 2014 MNRAS, 437, 475-488
37. Parsons, S. et al. 2013, MNRAS
38. Hardy A. et al. 2016 MNRAS, 459(4), 4518-4526
39. Shaw, J. & Lopez-Morales, M. 2007, ASP Conf. Series, 362
40. Coughlin, J. & Shaw, J. 2007, JSARA., I, 7

41  Wolf, M. et al. 2016, A&A, 587, A82
42  Zhang et al. 2018 Res. Astron. Astrophys.18 116
43  Schleicher, D.R. & Dreizler, S. 2014, A&A, 563, A61.
44  Parsons, S.G., Marsh, T., Bours, M.C, et al. 2014,  MNRAS, 438, L91.

*Appendix 1 ~ Evolutionary Scenarios for Post Common Envelope Binary Systems*

Various evolutionary scenarios have been proposed for PCEB systems, but a definitive mechanism remains to be found. Favoured models suggest that when the more massive primary evolves to the red giant phase it fills its Roche lobe and matter is transferred from the primary star to its smaller main sequence companion at a rate that cannot be accommodated by the smaller star.  This unstable mass transfer from the primary forms a common envelope that surrounds the helium burning core of the red giant primary and its smaller companion. As a consequence angular momentum is transferred from the binary system to the surrounding envelope bringing the binary pair closer together and resulting in a short binary period of typically between 2 and 3 hours.  Eventually the common envelope acquires sufficient angular momentum for it to be mostly ejected from the system, so creating a planetary nebula surrounding a detached binary system.

As the common envelope phase is of short duration the mass of the smaller companion remains substantially constant since it was unable to accommodate the mass transfer. The remaining mass of what is left of the red giant, the primary, is about equal to the mass of the core of the giant at the onset of mass transfer. The helium rich primary forms an sdB star, as is the case of HS0705+6700 and NSVS 14256825, and well on their way to becoming a white dwarf, for example NN Ser.  Eventually, the loss in angular momentum will bring the binary pair into close proximity, and so becoming a classic cataclysmic variable.

NSVS 01286630 is somewhat different being a system that has not yet filled its Roche lobes, but in time the more massive star will become a red giant and follow a similar path to other sdBs as described above.

Planet formation around PCEB systems remains an unanswered question and two scenarios are postulated.  The first generation hypothesis suggests planets are formed before the expulsion of the common envelope, so surviving this cataclysmic event, see for example Hardy et al. (2016) and by Parsons et al. (2014) using primarily the NN Ser system as a model.[38][44]

The second-generation hypothesis suggests planet formation occurs after the expulsion of the common envelope and from the remaining protoplanetary disc.  This hypothesis is strongly supported from simulations by Zorotovic & Schreiber (2013) and Schleicher & Dreizler (2014).[1] [43]

12***Appendix 2 ~ Tables 3 and 4***

| Telescope/Observatory | MPC Code | Reference (see Table 4) |
|---|---|---|
| 0.51m Gemini Univ. of Iowa | 857 | 1 |
| 0.51m SSON Australia | Q65 | 2 |
| 0.5m iTelescope T11 New Mexico | H06 | 3 |
| 0.32m iTelescope T18 Nerpio | I86 | 4 |
| 0.43m iTelescope T21 New Mexico | H06 | 5 |
| 0.61m iTelescope T24 California | U69 | 6 |
| 0.7m iTelescope T27 Warrumbungle | Q65 | 7 |
| 2m LCO Faulkes South | E10 | 8 |
| 2m LCO Faulkes North | F65 | 9 |
| 0.28m Greenmoor Obs.Oxfordshire | Z54 | 10 |
| 0.36m Astrognosis Obs. Essex | K01 | 11 |
| 0.2m Woodland Obs.W. Sussex | - | 12 |
| 0.23m Ham Obs. W. Sussex | - | 13 |

Table 3: Telescopes and Minor Planet Centre (MPC) codes used for the measurements reported in this paper.

| BJD | Error (days) | Cycle | Minima | Filter | Telescope |
|---|---|---|---|---|---|
| *HS0705+6700* | $T_0 = 2451822.76155$ | | | | |
| 2458050.889105 | 0.000036 | 65116 | I | Clear | 1 |
| 2458056.436624 | 0.000059 | 65174 | I | Sloan r' | 1 |
| 2458070.400908 | 0.000242 | 65320 | I | Johnson V | 12 |
| 2458144.336056 | 0.000030 | 66093 | I | Clear | 11 |
| 2458144.431452 | 0.000102 | 66094 | I | Clear | 11 |
| 2458144.527359 | 0.000041 | 66095 | I | Clear | 11 |
| 2458144.622960 | 0.000031 | 66096 | I | Clear | 11 |
| 2458161.552412 | 0.000025 | 66273 | I | Sloan r' | 11 |
| 2458162.413229 | 0.000046 | 66282 | I | Sloan r' | 11 |
| 2458162.509018 | 0.000083 | 66283 | I | Sloan r' | 11 |
| 2458212.436574 | 0.000126 | 66805 | I | Johnson V | 12 |
| 2458224.392311 | 0.000157 | 66930 | I | Johnson V | 12 |
| 2458226.496587 | 0.000275 | 66952 | I | Johnson V | 12 |
| 2458252.417033 | 0.000252 | 67223 | I | Johnson V | 12 |
| 2458258.442570 | 0.000203 | 67286 | I | Johnson V | 12 |
| 2458418.459648 | 0.000126 | 68959 | I | Johnson V | 12 |
| 2458425.441796 | 0.000155 | 69032 | I | Johnson V | 12 |
| 2458440.362812 | 0.000179 | 69188 | I | Johnson V | 12 |
| 2458440.458271 | 0.000206 | 69189 | I | Johnson V | 12 |
| 2458462.361467 | 0.000124 | 69418 | I | Johnson V | 12 |
| 2458485.412263 | 0.000084 | 69659 | I | Johnson V | 12 |
| 2458492.490163 | 0.000061 | 69733 | I | Johnson V | 12 |
| 2458512.480366 | 0.000236 | 69942 | I | Johnson V | 12 |
| 2458514.393247 | 0.000100 | 69962 | I | Johnson V | 12 |
| 2458514.488987 | 0.000107 | 69963 | I | Johnson V | 12 |
| 2458560.303749 | 0.000052 | 70442 | I | Clear | 11 |
| 2458560.494993 | 0.000027 | 70444 | I | Clear | 11 |
| 2458567.381801 | 0.000102 | 70516 | I | Johnson V | 12 |
| 2458567.477389 | 0.000186 | 70517 | I | Johnson V | 12 |
| 2458568.529338 | 0.000035 | 70528 | I | Clear | 11 |
| 2458569.390252 | 0.000067 | 70537 | I | Johnson V | 12 |
| 2458587.467371 | 0.000126 | 70726 | I | Johnson V | 12 |
| 2458593.397471 | 0.000098 | 70788 | I | Sloan r' | 10 |
| 2458603.440464 | 0.000077 | 70893 | I | Clear | 12 |
| 2458634.429954 | 0.000175 | 71217 | I | Clear | 12 |
| 2458708.460552 | 0.000166 | 71991 | I | Clear | 12 |
| 2458728.546356 | 0.000043 | 72201 | I | Clear | 13 |
| 2458728.546331 | 0.000037 | 72201 | I | Clear | 11 |
| 2458734.572115 | 0.000045 | 72264 | I | Clear | 13 |
| 2458740.502201 | 0.000102 | 72326 | I | Clear | 13 |
| *NN Ser* | $T_0 = 2447344.524368$ | | | | |
| 2457899.487998 | | 81142 | I | Clear | 1 |
| 2457908.073445 | | 81208 | I | Luminance | 7 |
| 2458176.949347 | 0.000040 | 83275 | I | Sloan g' | 1 |
| 2458216.884018 | 0.000117 | 83582 | I | Clear | 1 |
| 2458230.152189 | 0.000074 | 83684 | I | Luminance | 7 |
| 2458218.834981 | 0.000254 | 83597 | I | Clear | 1 |
| 2458220.916317 | 0.000138 | 83613 | I | Clear | 1 |
| 2458310.411610 | | 84301 | I | Clear | 10 |
| 2458311.452237 | 0.000135 | 84309 | I | Clear | 10 |
| 2458314.444288 | 0.000079 | 84332 | I | Clear | 10 |
| 2458377.663256 | | 84818 | I | Clear | 3 |
| 2458377.663086 | 0.000079 | 84818 | I | Clear | 6 |
| 2458603.092126 | 0.000017 | 86551 | I | Johnson V | 8 |
| 2458610.116434 | 0.000015 | 86605 | I | Johnson R | 8 |
| 2458666.961549 | 0.000013 | 87042 | I | Johnson R | 8 |
| 2458716.782250 | 0.000016 | 87425 | I | Johnson R | 9 |
| *NSVS 01286630* | $T_0 = 2454272.753960$ | | | | |
| 2458306.490912 | 0.000130 | 10506.5 | II | Johnson V | 12 |
| 2458315.514056 | 0.000110 | 10530 | I | Johnson V | 12 |
| 2458317.433500 | 0.000212 | 10535 | I | Johnson V | 12 |
| 2458322.424583 | 0.000150 | 10548 | I | Johnson V | 12 |
| 2458325.496040 | 0.000138 | 10556 | I | Johnson V | 12 |
| 2458332.406650 | 0.000150 | 10574 | I | Johnson V | 12 |
| 2458335.477738 | 0.000089 | 10582 | I | Johnson V | 12 |
| 2458345.459948 | 0.000124 | 10608 | I | Johnson V | 12 |
| 2458379.437658 | 0.000143 | 10696.5 | II | Johnson V | 12 |
| 2458380.397297 | 0.000049 | 10699 | I | Clear | 11 |
| 2458390.379420 | 0.000067 | 10725 | I | V Filter | 11 |
| 2458450.656220 | 0.000070 | 10882 | I | Sloan g' | 1 |
| 2458533.584079 | 0.000098 | 11098 | I | V Filter | 11 |
| 2458540.494919 | 0.000060 | 11116 | I | R Filter | 11 |
| 2458547.405424 | 0.000207 | 11134 | I | Sloan r' | 11 |
| 2458555.468052 | 0.000135 | 11154 | I | Sloan r' | 11 |
| 2458560.459039 | 0.000184 | 11168 | I | Johnson V | 12 |
| 2458608.450030 | 0.000065 | 11293 | I | Clear | 12 |
| 2458626.494596 | 0.000063 | 11340 | I | Clear | 12 |
| 2458679.476185 | 0.000180 | 11478 | I | Clear | 12 |
| 2458687.538711 | 0.000088 | 11499 | I | Clear | 12 |
| 2458728.618856 | 0.000071 | 11606 | I | IRB | 13 |
| *NSVS 14256825* | $T_0 = 2454274.209211$ | | | | |
| 2458017.436008 | 0.000043 | 20146 | I | Clear | 4 |
| 2458039.290180 | 0.000049 | 20344 | I | Sloan r' | 10 |
| 2458073.285319 | 0.000034 | 20652 | I | Luminance | 4 |
| 2458229.243900 | 0.000044 | 22065 | I | Johnson V | 2 |
| 2458264.343135 | | 22383 | I | Johnson V | 2 |
| 2458288.183832 | 0.000063 | 22599 | I | Johnson V | 2 |
| 2458288.294274 | 0.000139 | 22600 | I | Johnson V | 2 |
| 2458311.252050 | 0.000035 | 22808 | I | Johnson V | 2 |
| 2458341.163464 | 0.000074 | 23079 | I | Johnson V | 2 |
| 2458370.081490 | 0.000013 | 23341 | I | Clear | 2 |
| 2458395.357172 | 0.000057 | 23570 | I | Johnson V | 12 |
| 2458407.332810 | 0.000379 | 23678.5 | II | Johnson V | 12 |
| 2458411.361260 | 0.000163 | 23715 | I | Johnson V | 12 |
| 2458412.354860 | 0.000074 | 23724 | I | Johnson V | 12 |
| 2458440.610650 | 0.000240 | 23980 | I | V | 5 |
| 2458418.646110 | 0.000050 | 37549 | I | Sloan r' | 1 |
| 2458424.606320 | 0.000030 | 37603 | I | Sloan r' | 1 |
| 2458429.573220 | 0.000020 | 37648 | I | Sloan r' | 1 |
| 2458435.643740 | 0.000030 | 37703 | I | Sloan r' | 1 |
| 2458442.597240 | 0.000060 | 37766 | I | Sloan r' | 1 |
| 2458451.648070 | 0.000030 | 37848 | I | Sloan r' | 1 |
| 2458453.634790 | 0.000030 | 37866 | I | Sloan r' | 1 |
| 2458610.255774 | 0.000056 | 39285 | I | V | 2 |
| 2458662.131600 | 0.000020 | 39755 | I | V | 2 |
| 2458692.043017 | 0.000223 | 40026 | I | V | 2 |
| 2458692.153239 | 0.000067 | 40027 | I | V | 2 |
| 2458721.402501 | 0.000022 | 40292 | I | Clear | 12 |
| 2458722.395865 | 0.000056 | 40301 | I | IRB | 13 |
| 2458726.369288 | 0.000026 | 40337 | I | IRB | 13 |
| 2458728.466423 | 0.000016 | 40356 | I | IRB | 13 |

Table 4: Eclipse time minima observed between 2017 May and 2019 September. The reference epoch is noted for each binary system in each section header. See Table 3 for telescope reference.